\documentclass[runningheads,psfig]{cl2emult}

\usepackage{makeidx}  
\usepackage{psfig}
\usepackage{subeqnar} 
\usepackage{multicol} 
\usepackage{cropmark} 
\usepackage{eso}      
\makeindex            

\begin{document}
\title*{Radio Cores in Low-Luminosity AGN: ADAFs or Jets?}
\toctitle{Radio Cores in Low-Luminosity AGN and Binary Black Holes: ADAFs or Jets?
}
%
%
\titlerunning{Radio Cores in LLAGN: ADAF or Jet?}
%
\author{Heino Falcke\inst{1} \and Neil M. Nagar\inst{2} \and Andrew S. Wilson\inst{2} \and Luis C. Ho\inst{3} \and Jim S. Ulvestad\inst{4}}
\authorrunning{Falcke et al.}
%
%
\institute{Max-Planck-Institut f\"ur Radioastronomie, Auf dem H\"ugel 69,
D-53121 Bonn, Germany (hfalcke@mpifr-bonn.mpg.de)
\and
Dept. of Astronomy, University of Maryland, College Park, MD
20742-2421, USA (wilson,neil@astro.umd.edu)
\and 
Carnegie Observatories, 813 Santa Barbara Street, Pasadena, CA 91101,
USA (lho@ociw.edu)
\and
NRAO, P.O. Box O, 1003 Lopezville Road, Socorro, NM 87801 (julvesta@aoc.nrao.edu)
}

\maketitle              

\begin{abstract}
We have surveyed two large samples of nearby low-luminosity AGN with
the VLA to search for flat-spectrum radio cores, similar to Sgr A* in
the Galactic Center. Roughly one third of all galaxies are detected
(roughly one half if {H}{\sc II} transition objects are excluded from
the sample), many of which have compact radio cores. Follow-up
observations with the VLBA have confirmed that these cores are
non-thermal in origin, with brightness temperatures of $\ge10^8$
K. The brightest of these are resolved into linear structures. The
radio spectral indices of the cores are quite flat ($\alpha\sim0$),
with no evidence for the highly inverted radio cores predicted in the
ADAF model. Spectrum and morphology of the compact radio emission is
typical for radio jets seen also in more luminous AGN. The
emission-line luminosity seems to be correlated with the radio core
flux. Together with the VLBI observations this suggests that optical
and radio emission in at least half the low-luminosity Seyferts and
LINERs are black hole powered. We find only a weak correlation between
bulge luminosity and radio flux and an apparently different efficiency
between elliptical and spiral galaxies for producing radio emission at
a given optical luminosity.
\end{abstract}

\section{Introduction}
\footnotetext[5]{In: ``Black Holes in Binaries and Galactic Nuclei'', ESO workshop,
eds. L. Kaper, E.P.J. van den Heuvel, P.A. Woudt, Springer Verlag}
What powers the nuclei of nearby galaxies? Many of them show evidence
for emission-lines similar to those seen in active galactic nuclei
(AGN) but on a much lower level (Ho et al. 1997a) --- therefore they
are called low-luminosity AGN (LLAGN). In some cases broad lines are
seen and hence one infers the presence of a central black hole (Ho et
al. 1997b).  In most cases, however, even a moderate starburst might be 
able to explain the observed optical spectra (Alonso-Herrero et al. 1999),
especially those residing in LINER galaxies (Heckman et al. 1983).

Another method to identify the nature of the activity is to search for
compact, flat-spectrum radio cores with high brightness temperatures,
since this is a typical feature of many AGN and cannot be explained by
star formation. For LLAGN the nature of these radio cores is largely
unclear. It has been proposed that the compact radio emission could be
produced either by emission from an Advection Dominated Accretion Flow
(ADAF; e.g. Narayan et al. 1998) or from scaled-down AGN jets
(Falcke \& Biermann 1996; 1999).

We have therefore performed a VLA survey of two samples of nearby
galaxies with optical emission-lines to identify such compact radio
cores. Follow-up observations with the VLBA of these cores have been
made that shed further light on their nature.

\section{Samples and Observations}
The first sample we observed consisted of 48 galaxies with mainly
LINER-like emission spectra that were part of ongoing studies at other
wavelengths. In a second project we expanded this sample to a
distance-limited sample of galaxies with emission-lines within 19 Mpc.

Both samples were observed with the VLA in its largest configuration
at 15 GHz. In the final data reduction we reached a 10 $\sigma$
detection limit of $\sim1.1$ mJy. The resolution was about
0.15$^{\prime\prime}$ which corresponds to a linear scale of 14 pc for
a galaxy at a distance of 19 Mpc. All sources which were detected with
compact emission above 3 mJy in either sample were then observed with
the VLBA at 5 GHz with a resolution of 2.5 mas ($\sim0.2$ pc at 19 Mpc
distance) and a detection limit around 2 mJy.

\section{Results}
We are going to restrict the following discussion to the detection of
compact core emission. The detection rate in our first LLAGN sample was
35\% (17 of 48), higher than similar deep surveys of normal galaxies
(Wrobel \& Heeschen 1991). Only two sources had steep spectra and only
one out of eighteen sources with optical classification as transition
sources (Ho et al. 1997a) was detected. The other detections are all in
LLAGN with either Seyfert or LINER spectra. This is confirmed by the
results of our distance-limited survey: 44\% of LLAGN with Seyfert or
LINER spectra have compact cores, but only 12\% of transition objects
do.

\begin{figure}
\centering
\noindent
\psfig{figure=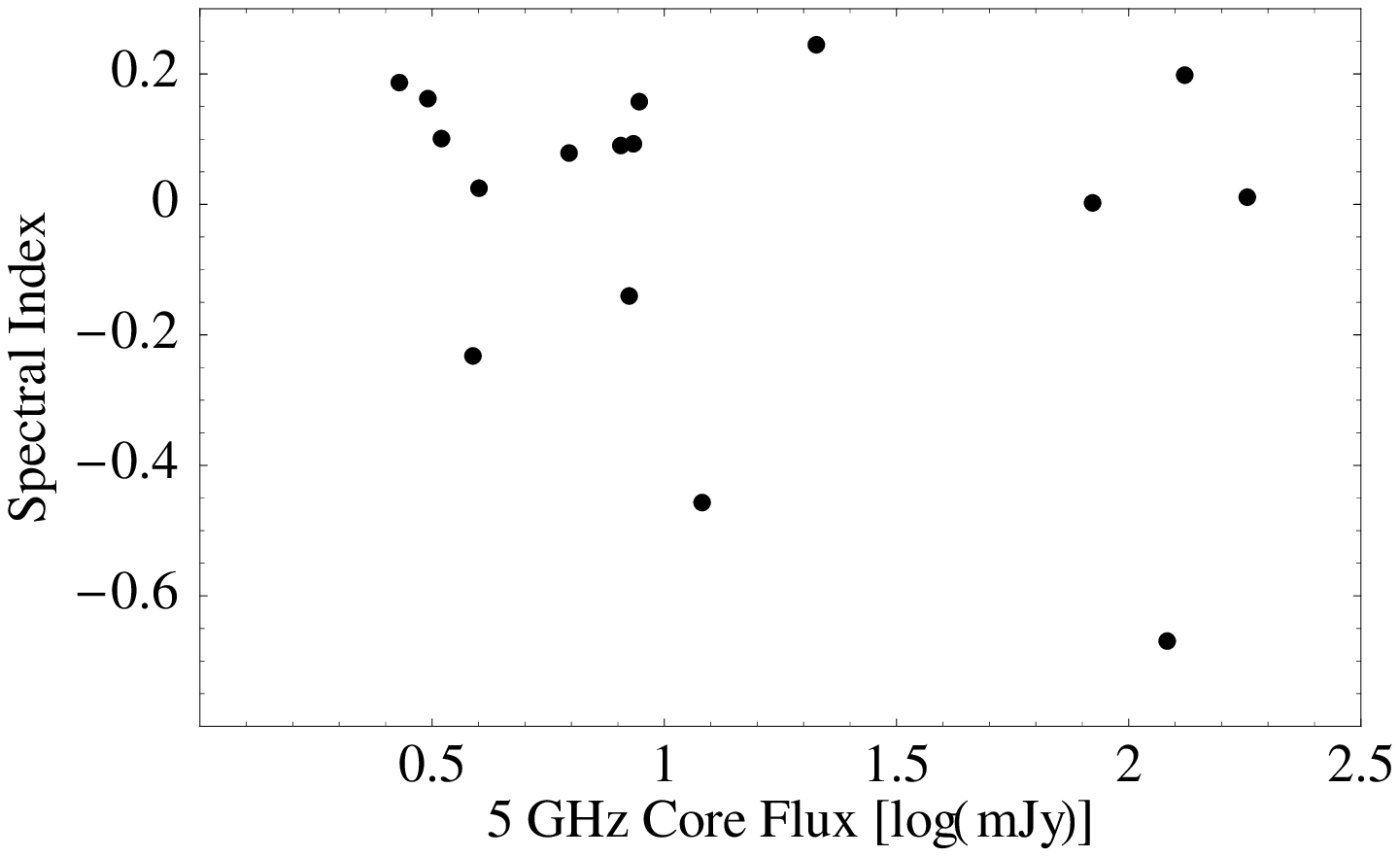,width=0.49\textwidth,bbllx=3.1cm,bblly=18.1cm,bburx=17.8cm,bbury=27cm,clip=}\psfig{figure=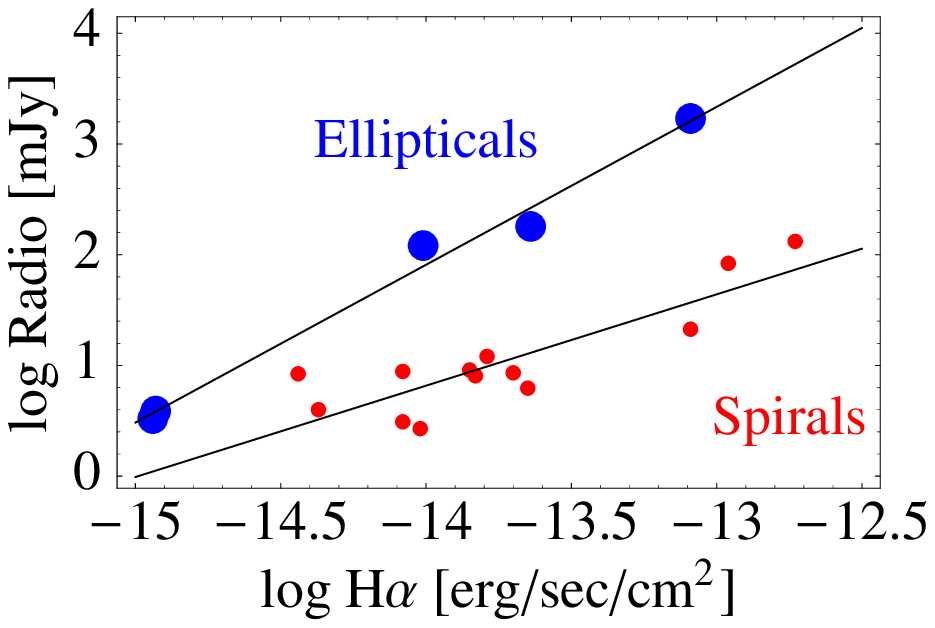,width=0.49\textwidth,bbllx=3.8cm,bblly=20.8cm,bburx=13.8cm,bbury=27cm,clip=}
\caption[]{Left: Spectral indices of LLAGN in our sample with $S_{\rm 15 GHz}>3$
 mJy between 5 GHz (VLBI) and 15 GHz (VLA) as a function of radio core
flux at 5 GHz. Right: $S_{\rm 15 GHz}$ plotted versus narrow H$\alpha$
flux for the same sample; ellipticals and spirals are distinguished by
big and small dots respectively.}
\label{ha}
\end{figure}

These results suggest that galaxies with Seyfert and LINER spectra are
black hole powered, while transition objects are dominated by star
formation. The evidence for black hole powered engines is further
strengthened by our VLBA results. Even though our detection limit was
close to our selection threshold, 19 out of 20 galaxies\footnote{This
includes M81 and M87 which are part of the sample but have well known
radio cores and were not observed by us.} showed compact radio emission
with brightness temperatures of the order $T_{\rm B}\ge10^8$ K. The
one non-detected source had a steep-spectrum and hence is the
exception which confirms the rule that also in LLAGN flat-spectrum
radio cores are a sign of high-$T_{\rm B}$ AGNs. We find that the six
brightest sources in our VLBI sample all show typical core-jet
structures. The fainter cores probably have too low dynamic range and
signal-to-noise to show any significant extended structure. Figure
\ref{ha} (left panel) shows the distribution of spectral indices
between our total 6 cm (VLBA) and 15 GHz (VLA) flux densities
($S_\nu\propto\nu^\alpha$). Even though comparing VLBA with VLA fluxes
and our selection at 15 GHz is biased towards highly inverted spectra,
none of the spectral indices has $\alpha>0.25$, in conflict with the
prediction of the ADAF model (e.g. Yi \& Boughn 1998) but quite
consistent with the predictions of jet models (Falcke \& Biermann
1999). The average is $\left<\alpha\right>=0.0$.

For the VLBI-sample, i.e. the well-detected cores above 3 mJy, for
which we have basically established that the radio emission is
AGN-related, we also looked at correlations between radio,
emission-line, and bulge luminosities. Figure \ref{ha} (right panel)
shows that there is a trend for galaxies with higher H$\alpha$
emission to have more luminous radio cores. Interestingly, elliptical
and spiral host galaxies are offset from each other. Does this reflect
a radio-loud/radio-quiet dichotomy for LLAGN?

\begin{figure}
\centering
\noindent
\psfig{figure=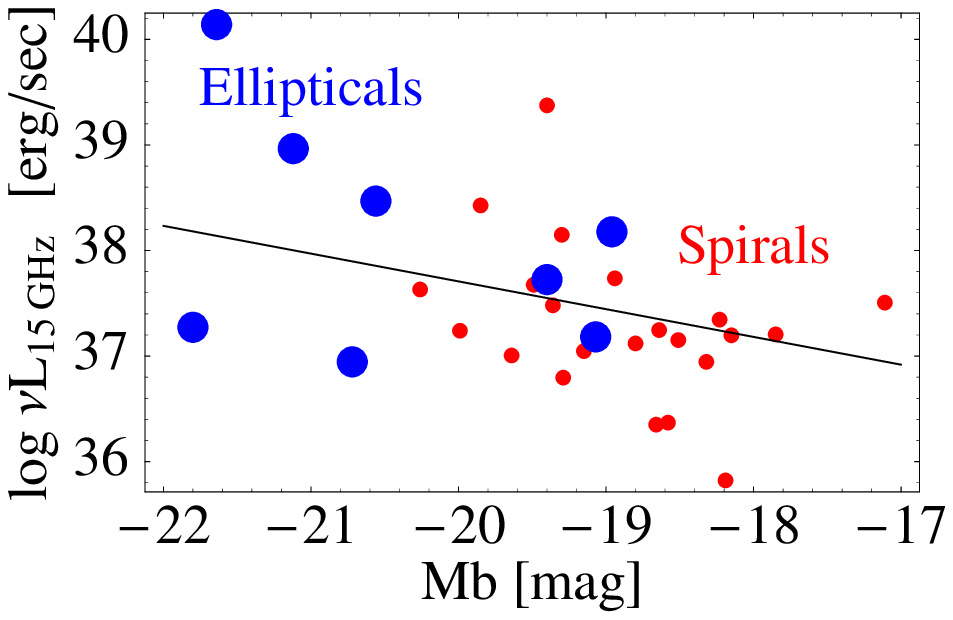,width=0.475\textwidth,bbllx=3.8cm,bblly=20.8cm,bburx=13.4cm,bbury=27cm,clip=}\psfig{figure=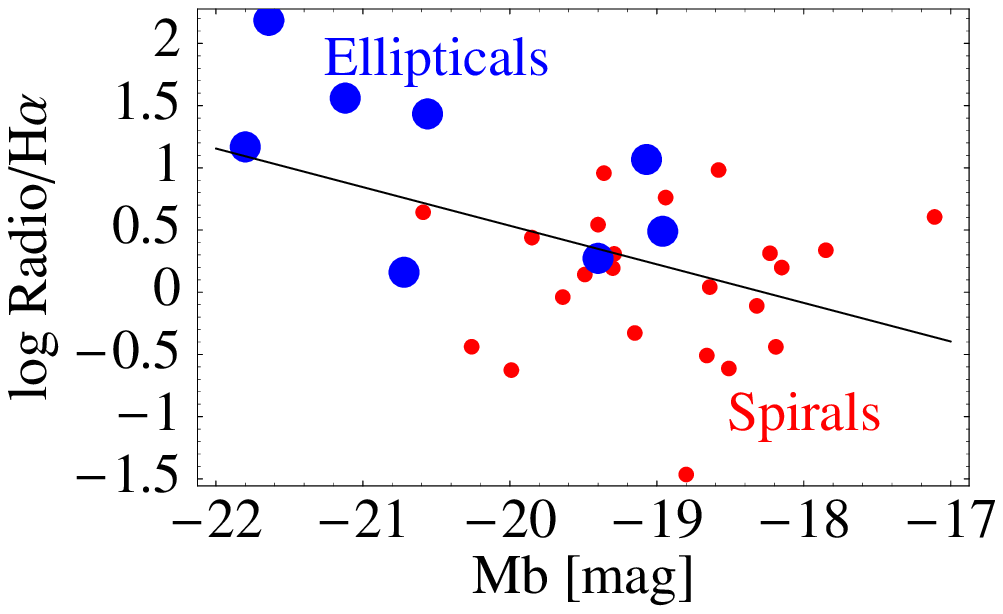,width=0.505\textwidth,bbllx=3.5cm,bblly=20.9cm,bburx=13.7cm,bbury=27cm,clip=}
\caption[]{Left: Radio luminosity ($\nu L_\nu$) at 15 GHz of LLAGN in our sample 
with $S_{\rm 15 GHz}>1.5$ mJy as a function of blue bulge
magnitude. Right: Ratio between 15 GHz radio core and narrow H$\alpha$
flux as a function of blue bulge magnitude in the same sample.
Ellipticals and spirals are distinguished by big and small dots
respectively.}
\label{Mb}
\end{figure}

However, there is another important factor:
the galaxy bulge luminosity. We do see a weak trend for the radio
luminosity to be related to bulge luminosity; also the ratio between
radio and H$\alpha$ luminosity tends to increase with increasing bulge
luminosity. Hence, galaxies apparently become more efficient in
producing radio emission relative to H$\alpha$ in bigger bulges. This
also holds if we look at the enitre VLA detected sample
(Fig.~\ref{Mb}). Whether this is due to increasing obscuration,
effects intrinsic to the AGN, or a selection effect is unclear.  Since
ellipticals and spirals in our sample are nicely separated between the
top and bottom end of the bulge luminosity distribution, an apparent
dichotomy in Fig.~\ref{ha} is a natural consequence of this trend.

\section{Discussion \& Summary}
We find that at least 40\% of optically selected LLAGN with Seyfert
and LINER spectra have compact radio cores. VLBI observations show
that these cores are similar to radio jets in more luminous AGN with
high brightness temperatures, jet-like structures, and flat radio
spectra (e.g.~Falcke \& Biermann 1996; 1999). The radio emission seems
to be related to the luminosity of the emission-line gas and hence
both are probably powered by genuine AGN operating at low powers. We
found no evidence for high frequency components with highly inverted
spectra predicted in ADAF models. Hence, for these models one should
probably not include radio fluxes in broad-band spectral fits. We also
find only a weak correlation between radio and bulge luminosity.
Together with the radio-H$\alpha$ correlation this makes it very
unlikely that the black hole mass could be reliably determined from
the radio data---in contrast to what is occasionally suggested.

\clearpage
\addcontentsline{toc}{section}{Index}
\flushbottom
\printindex


\begin{thebibliography}{7}
%
\addcontentsline{toc}{section}{References}
\bibitem{}Alonso-Herrero, A., Rieke, M.J., Rieke, G.H., Shields, J.C. 1999, ApJ, in press [astro-ph/9909316]
\bibitem{}{Falcke, H., \& Biermann, P.L. 1996, A\&A 308, 321}
\bibitem{}{Falcke, H., \& Biermann, P.L. 1999, A\&A 342, 49}
\bibitem{}{Heckman, T. M., Van Breugel, W., Miley, G. K., Butcher, H. R.
1983, AJ, 88, 1077}
\bibitem{}{Ho, L.~C., Filippenko, A.~V., \& Sargent, W.~L.~W. 1997a, ApJ, 487, 568}
\bibitem{}{Ho, L.~C., Filippenko, A.~V., \& Sargent, W.~L.~W. 1997b, ApJS, 112, 391}
\bibitem{}{Narayan, R., Mahadevan, R., Grindlay,  J.~E., Popham, R.G., \& Gammie, C. ~1998, ApJ 492, 554}
\bibitem{}{Wrobel, J. M., \& Heeschen, D. S. 1991, AJ, 101, 148}
\bibitem{}{Yi, I., \& Boughn, S. P. 1998, ApJ, 499, 198}
\end{thebibliography}
\end{document}